\documentclass[12pt]{article}
\usepackage{amsfonts,amsmath,epsfig,psfrag}
\usepackage{amsmath,amsfonts,amssymb}

\def\ii{{\it i}}

\def\r{{ R}}
\def\t{{ T}}
\def\o{{\bf 0}}
\def\q{{ Q}}
\def\i{{ I}}
\def\bm#1{\mbox{\boldmath{$#1$}}}

\newcommand{\be}{\begin{equation}}
\newcommand{\en}{\end{equation}}

\newcommand{\la}{\label}
\def\rr#1{(\ref{#1})}

\newcommand{\filler}{\hspace*{\fill}}

\begin{document}

\title{The speed of interfacial waves  \\
polarized in a symmetry plane}

\author{
  \filler Michel Destrade$^{a,*}$, YiBin. Fu$^{b}$, \filler\\
  {\it\normalsize \filler $^a$Laboratoire de Mod\'elisation en M\'ecanique,}
         \filler\\
  {\it\normalsize \filler CNRS (UMR7607), Universit\'e Pierre et Marie Curie,}
          \filler\\
  {\it\normalsize \filler Case 162, 4 Place Jussieu, 75252 Paris Cedex 05,
       France}\filler\\[9pt]
  {\it\normalsize \filler $^b$Department of Mathematics, Keele University,
  Staffordshire ST5 5BG, UK}\filler
  }
\date{}
\maketitle

\begin{abstract}
The surface-impedance matrix method is used to study interfacial
waves polarized in a plane of symmetry of anisotropic elastic
materials. Although the corresponding Stroh polynomial is a
quartic, it turns out to be analytically solvable in quite a
simple manner. A specific application of the result concerns the
calculation of the speed of a Stoneley wave, polarized in the
common symmetry plane of two rigidly bonded anisotropic solids.
The corresponding algorithm is robust, easy to implement, and
gives directly the speed (when the wave exists) for any
orientation of the interface plane, normal to the common symmetry
plane. Through the examples of the couples (Aluminum)-(Tungsten)
and (Carbon/epoxy)-(Douglas pine), some general features of a
Stoneley wave speed are verified: the wave does not always exist;
it is faster than the slowest Rayleigh wave associated with the
separated half-spaces.
\end{abstract}


\newpage 

\section{Introduction}

The semiconductor industry makes great use of wafer bonding, a
process which allows two different materials to be rigidly and
permanently bonded along a plane interface, thus producing a
composite bi-material \cite{GoTo98}. Worldwide, there are now
several hundreds of wafer bonding patents deposited yearly
\cite{AlGo03}. A similar process, fusion bonding, is used by the
polymer industry to bring together two parts of different solid
polymers, thus enabling the manufacture of a heterogeneous
bi-material with specific properties \cite{Harp96, AgYe02}. In the
first case, wafers of two different crystals are stuck together
through van der Waals forces, after their surface has been
mirror-polished; in the second context, a fusion process takes
place at the interface, followed by a cooling and consolidating
period. Whichever the process, it seems important to be able to
inspect the strength of the bonding, possibly through
non-destructive ultrasonic evaluation. This is where the study of
Stoneley waves (rigid contact) and of slip waves (sliding contact)
is relevant (see Rokhlin et al. \cite{RoHR80, RoHR81} or Lee and
Corbly \cite{LeCo77}). In the continuum mechanics literature,
great contributions can be found on the theoretical apprehension
of these waves. In particular, Barnett et al. \cite{BLGM85} for
Stoneley waves and Barnett et al. \cite{BaGL88} for slip waves
have provided a rigorous and elegant corpus of results for their
possible existence and uniqueness, based on the Stroh formalism
\cite{Stro62, Ting96}. However very few simple numerical
``recipes'' exist to compute the speed (and then the attenuation
factors, partial modes, and profiles) of these waves when they
exist.

When the two materials have at least orthorhombic symmetry and
their crystallographic axes are aligned, the
analysis can be conducted in explicit form as is best summarized
in the article by Chevalier et al. \cite{ChLM91}.
This explicit analysis
is possible because for waves proportional to $\exp ik(x_1 + px_2
-vt)$ where $k$, $p$, $v$ are the respective wave number, attenuation
factor, and speed of the wave, and $x_1$, $x_2$
(aligned with two common crystallographic axes)
are the respective directions of propagation and attenuation,
the equations of motion lead to a \textit{propagation condition}
which is a \textit{quadratic} in $p^2$; then the relevant roots
$p$ can be found in terms of the stiffnesses $C_{ij}$, $C^*_{ij}$
for each half-space, of the mass densities $\rho$, $\rho^*$, and
of the speed $v$. After construction of the general solution to
the equations of motion (a linear superposition of the partial
modes), the boundary condition at the interface (rigid or sliding
contact) yields the \textit{secular equation}, of which $v$ is a
root. If the crystallographic axes of the two orthorhombic
materials do not coincide, or when at least one material is
monoclinic or triclinic, then the situation becomes much more
intricate. In the very particular case where the bi-material is
made of the \textit{same} material above and below the plane
interface, with crystallographic axes \textit{symmetrically}
misoriented, a Stoneley wave can be found along the bisectrix of
the misorientation angle \cite{LiMu70, MoTW98, Dest04}.
Otherwise, analytical methods fail because
in general the propagation condition is a \textit{sextic} in $p$,
unsolvable \cite{Head79} in the Galois sense.

Now consider the case of a bi-material made of two
\textit{different} anisotropic half-spaces with a common symmetry
plane $Z=0$, orthogonal to the interface, and with only one common
crystallographic axis (the $Z$ axis), see Figure 1.
\begin{figure}[!t]
\epsfig{figure=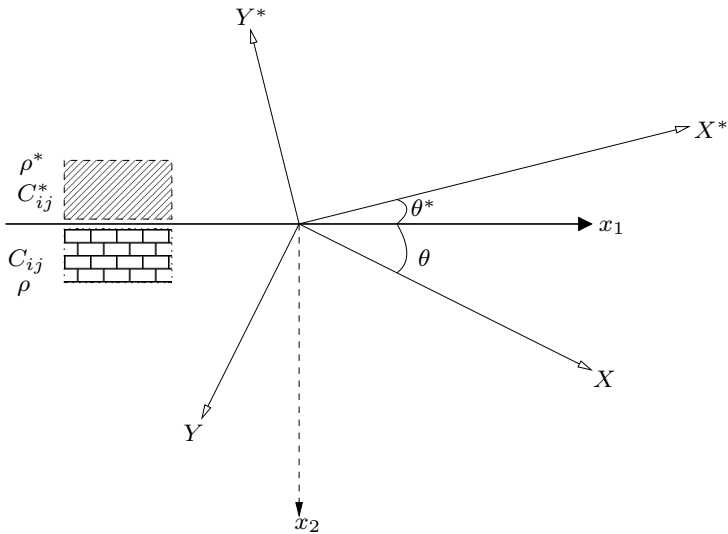,width=.7\textwidth}
 \renewcommand{\figurename}{Fig.}
\caption{Plane
interface (at $x_2 = 0$) between two different anisotropic
materials with a common symmetry plane (at $Z=0$).}
\end{figure}
Then the in-plane strain decouples from the anti-plane strain
\cite{Stro62} and in each half-space, the propagation condition
factorizes into the product of a quadratic in $p$ (associated
with the anti-plane strain) and a \textit{quartic} in $p$
(associated with the in-plane strain). Solving analytically this
quartic and identifying the two qualifying roots unambiguously in
order to be able to write down the boundary condition may have
seemed a formidable task. However, it was recently shown in Fu
\cite{Fu05} that such a quartic can in fact be solved and the two
qualifying roots identified in quite a simple manner, using
results from algebra. Once the roots $p$ are known, the
complete resolution of the problem flows out naturally because the
surface-impedance matrices $M(v)$ and $M^*(v)$ for each region are
now known explicitly. This knowledge gives in turn an explicit
secular equation, in the form $f(v) = 0$ for some function $f$.
For Stoneley waves, $f$ is a monotone decreasing function of $v$,
whose \textit{only} zero, when it exists, is the wave speed. This
latter property is worth emphasizing: therein lies the superiority
of the explicit surface-impedance matrix method above others based
on algebraic manipulations which ultimately lead to a multitude of
secular equations \cite{Ting02a, Ting02b} and/or of spurious roots
\cite{Curr79, TaCu81, Tazi89, Mozh95, MoTW98, Dest01, Dest03a,
Dest03b}.

This contribution builds on recent advances by Fu and
Mielke \cite{FuMi02}, Mielke and Fu \cite{MiFu03}, and Fu \cite{Fu05},
themselves resting on
the major works by Barnett and Lothe \cite{BaLo85},
Chadwick and Smith \cite{ChSm77}, Ting \cite{Ting96},
and Mielke and Sprenger \cite{MiSp98}.
%We note that in the papers by Barnett et al. \cite{BLGM85, BaGL88}
%the authors proposed a two-to-three step procedure: (i) implement
%an algorithm which computes numerically the ``limiting speed''
%(defined later in this paper); (ii) use that value to perform a
%numerical test; if the test is negative, then the Stoneley wave
%does not exist; if the test is positive then (iii) implement
%another iterative algorithm to compute the speed of the Stoneley
%wave. Their method is valid even when the two crystals do not
%share a common symmetry plane, orthogonal to the interface plane.
%Its implementation, however, requires a detailed study and an
%in-depth understanding of the integral formalism \cite{BaLo74}.
%Here we present a single, direct algorithm for Stoneley waves
%polarized in a symmetry plane, which is a test in itself: if it
%yields a secular equation $f(v)=0$ with no root, then the Stoneley
%wave does not exist; if it yields a secular equation $f(v)=0$ with
%a root, then the Stoneley wave exists, the root is unique and is
%the speed of the wave.
We aim at keeping the algebra to a minimum and at delivering a
numerical recipe giving the wave speed in a robust manner. The
reader who is keen on implementing such procedures may skip the
next two sections to jump directly to Section 4, where they are
summarized and presented for the Rayleigh wave speed and for the
Stoneley wave speed.  Section 2 recapitulates the basic equations
of motion and presents the surface-impedance matrices. In Section
3, the quartic evoked above is derived, and then explicitly solved
for the roots which allow for a localization of the wave near the
interface. Finally, using the ``recipes'' of Section 4, examples
of Rayleigh and Stoneley wave speeds computations are presented in
Section 5, and the connection is made with some numerical results
of Chevalier et al. \cite{ChLM91}.

%Although not conducted here, the application of the procedure to
%other contexts is straightforward (slip waves, Scholte waves,
%Bleustein-Gulyaev waves, Maerfeld-Tournois waves, etc.). In
%effect, the knowledge of the surface-impedance matrix permits the
%apprehension and the resolution of a multitude of problems in
%elastodynamics and elastostatics, including \ldots

\section{Governing equations}

Consider a bi-material made of two distinct anisotropic materials
with a common symmetry plane, bonded rigidly along a plane
interface, $x_2=0$ say. Let $\rho$ and $C_{ijks}$ be the mass
density and elastic stiffnesses of the body below ($x_2 \ge 0$)
and $\rho^*$ and $C^*_{ijks}$ be those of the body above ($x_2 \le
0$). The $C_{ijks}$ and $C^*_{ijks}$ are assumed to satisfy the
symmetry relations \be C_{ijks} = C_{ksij} = C_{jiks}, \quad
\text{and} \quad C^*_{ijks} = C^*_{ksij} = C^*_{jiks}, \la{sym}
\en and the strong convexity conditions \be C_{ijks} \xi_{ij}
\xi_{ks} > 0,  \quad C^*_{ijks} \xi_{ij} \xi_{ks}
> 0, \quad \forall \hbox{ non-zero real symmetric tensors
}\bm{\xi}. \la{SC} \en The strong ellipticity conditions are given
by \be C_{ijks} \eta_{i}\eta_k \gamma_{j}\gamma_s > 0, \quad C^*_{ijks} \eta_{i}\eta_k \gamma_{j}\gamma_s > 0,
\quad \forall \hbox{ non-zero real vectors } \bm{\eta} \text{ and
} \bm{\gamma}, \la{SE} \en and are implied by the strong convexity
conditions \rr{SC}. Let $XYZ$ and $X^*Y^*Z$ be along the
crystallographic axes of each material; the $X$ axis ($X^*$ axis)
makes an angle $\theta$ ($\theta^*$) with the interface, and $Z$
is normal to their common symmetry plane and to $x_2$. Finally,
let $x_1$ be an axis such that $x_1x_2Z$ is a rectangular
coordinate system. See Figure 1.

In this context, an interfacial wave is a two-component
\cite{Stro62} inhomogeneous plane wave, whose propagation is
governed by the equations of motion for the mechanical
displacement ${\bm u}(x_1,x_2,t) = [u_1,u_2]^T$, \be C_{ijks}
u_{k,sj} = \rho \ddot{u}_i \quad (x_2 \ge 0), \quad C^*_{ijks}
u_{k,sj} = \rho^* \ddot{u}_i \quad (x_2 \le 0), \la{motion} \en
and which decays away from the interface, \be {\bm u} \rightarrow
\mathbf{0} \quad \text{as} \quad x_2 \rightarrow \pm \infty.
\la{decay} \en Here and henceforward, a comma denotes
differentiation with respect to spatial coordinates and a dot
denotes material time derivative. Since the surfaces of the upper
half-space and of the lower half-space have unit normals
($\delta_{2i}$) and $(-\delta_{2i})$, respectively, the traction
vectors on these two surfaces are \be t_i = - C_{i2ks} u_{k,s},
\quad t^*_i = C^*_{i2ks} u_{k,s}, \quad i=1,2, \la{ti2} \en and by
\rr{decay}, they also decay, \be t_i  \rightarrow 0 \quad
\text{as} \quad x_2 \rightarrow + \infty, \quad t^*_i  \rightarrow
0 \quad \text{as} \quad x_2 \rightarrow - \infty. \en Without loss
of generality, the interfacial wave is assumed to propagate along
the $x_1$-direction and to have unit wave number, so that \be {\bm
u} = {\bm z}(\ii x_2) {\rm e}^{\ii (x_1 - v t)}
 + {\rm c.c.},
 \la{u}
\en
where $v$ is the propagation speed and ``c.c.'' denotes
the complex conjugate of the preceding term.

Substituting \rr{u} into \rr{motion} and \rr{ti2}$_1$ gives
\be
\t {\bm z}'' + (\r+\r^T) {\bm z}' +(\q-\rho v^2 \i) {\bm z}
=\o, \quad x_2 \ge 0,
\la{2.9}
\en
and
\be
{\bm t} = -\ii {\bm l}(\ii x_2) {\rm e}^{\ii (x_1-v t)}
          + {\rm c.c.},
\la{2.10}
\en
where
\be
{\bm l}=\t {\bm z}'+ \r^T {\bm z},
\la{2.11}
\en
a prime signifies differentiation with respect to the argument
$\ii x_2$, and the $2 \times 2$ matrices $\t, \r, \q$ are defined
by their components,
\be
T_{ik} = C_{i2k2}, \quad
R_{ik} = C_{i1k2}, \quad
Q_{ik} = C_{i1k1}.
\la{2.12}
\en
Identical results apply for the upper half-space $x_2 \le 0$,
with each quantity replaced by its starred counterpart,
except that ${\bm t}^*$ is given by
\be
{\bm t}^* = \ii {\bm l}^*(\ii x_2) {\rm e}^{\ii (x_1-v t)}
          + {\rm c.c.}
\en
Note that satisfaction of the strong ellipticity conditions
\rr{SE} ensures that $T$, $T^*$, and $Q$, $Q^*$ are
all positive definite and hence invertible.

The {\it surface-impedance matrices} $M(v)$ and $M^*(v)$ are
defined by \be -\ii {\bm l}(0) = M(v) {\bm z}(0), \quad \ii {\bm
l}^*(0) = M^*(v) {\bm z}^*(0). \la{M} \en In the Stroh
\cite{Stro62} formulation, the second-order differential equation
\rr{2.9} is written as a system of first-order differential
equations for the variables ${\bm z}$ and ${\bm l}$. Thus, for the
lower half-space (and similarly for the upper half-space), \be
{\bm \xi}' = N {\bm \xi}, \quad \text{where} \quad {\bm \xi}=
\begin{bmatrix} {\bm z}\\ {\bm l}\end{bmatrix},
\quad
N = \begin{bmatrix}
      N_1 & N_2 \\
      N_3 + \rho v^2 I & N_1^T
    \end{bmatrix},
\la{2.14}
\en
and
\be
 N_1 = -T^{-1} R^T, \quad
 N_2 = T^{-1}, \quad
 N_3 = R T^{-1} R^T-Q.
\la{2.15}
\en
Here $I$ is the $2 \times 2$ identity matrix.

An important property of the surface impedance matrix is that it
is independent of the depth, that is
\begin{equation} \label{Mindependent}
  -{\bm l}(ix_2) = M {\bm z}(ix_2),
\end{equation}
see Ingebrigsten and Tonning \cite{InTo69}.
On substituting \eqref{Mindependent} into \eqref{2.14} and
eliminating ${\bm z}'$, we obtain
\begin{equation}
  \left\{ (M - iR)T^{-1} (M+iR^T) - Q + \rho v^2 \right\}
        {\bm z}(ix_2) = \mathbf{0}.
\end{equation}
Since ${\bm z}(ix_2)$ is arbitrary, it then follows that
\begin{equation} \label{Ricatti}
  (M-iR)T^{-1}(M+iR^T) - Q - \rho v^2 I = 0.
\end{equation}
This simple matrix equation satisfied by $M$ was seemingly first
derived by Biryukov \cite{Biry85}, and later rediscovered by Fu
and Mielke \cite{FuMi02} where it was shown how this equation
could be exploited to compute the surface speed.

The general solution of \rr{2.14} can be constructed by first
looking for a partial mode solution of the form \be {\bm \xi}(\ii x_2) = {\rm
e}^{\ii p x_2} {\bm \zeta}, \la{2.16} \en say, where $p$ is a
constant scalar  (attenuation factor) and ${\bm \zeta}$ is a
constant vector to be determined. Substituting \rr{2.16} into
\rr{2.14} leads to the eigenvalue problem $(N-p I_{4 \times 4})
{\bm \zeta} = {\bm 0}$. Under the assumption of strong
ellipticity, the eigenvalues of $N$ appear as two pairs of complex
conjugates when $v=0$ and they remain so until $v = v_c$, where
$v_c$ is referred to as the {\it limiting speed}. In this paper we
are only concerned with the subsonic case, for which $0 \le v <
v_c$. The solution \rr{2.16} decays as $x_2 \rightarrow \infty$
only if the imaginary part of $p$ is positive. Thus, denoting by
$p_1$, $p_2$ the two eigenvalues of $N$ with positive imaginary
parts, and by ${\bm \zeta}^{(1)}$, ${\bm \zeta}^{(2)}$ the
corresponding eigenvectors, a general decaying solution is given
by \be {\bm \xi}(\ii x_2) =
 c_1 {\rm e}^{\ii p_1 x_2}{\bm \zeta}^{(1)}
 +  c_2 {\rm e}^{\ii p_2 x_2}{\bm \zeta}^{(2)},
 \la{2.17}
\en
where $c_1$, $c_2$ are disposable constants.
Hence at the interface,
\be
{\bm \xi}(0) =
 c_1 {\bm \zeta}^{(1)} + c_2 {\bm \zeta}^{(2)}
 = \begin{bmatrix} A \\ B \end{bmatrix} {\bm c},
\la{2.18}
\en
where the $2 \times 2$ matrices $A$, $B$ and the column vector
${\bm c}$ are defined by
\be
\begin{bmatrix} A \\ B \end{bmatrix}
 = \left[{\bm \zeta}^{(1)} |
             {\bm \zeta}^{(2)} \right],
\quad {\bm c} = \begin{bmatrix} c_1 \\ c_2 \end{bmatrix}.
\la{2.19} \en It follows from \rr{2.18} that \be -\ii {\bm l}(0)
 = -\ii B {\bm c}
 = -\ii B A^{-1} {\bm z}(0),
 \la{2.20}
\en and so from \rr{M} that \be M(v) = -\ii BA^{-1}, \la{2.21} \en
which is a well-known representation of the surface-impedance
matrix. As is recalled later in the paper, the surface-impedance
matrices are crucial to the determination of the interfacial wave
speed. In fact, if their explicit expressions are found, then an
exact secular equation (an equation of which the wave speed is the
only zero) is also found explicitly.

%%%%%%%%%%%%%%%%%%%%%%%%
\section{Explicit expressions for the surface-im\-pe\-dan\-ce matrices}
%%%%%%%%%%%%%%%%%%%%%%%%

Here it is seen that the relevant roots to the characteristic
equation $\text{det}(N-p I_{4 \times 4}) = 0$ can be obtained
explicitly, without the uncertainty which one encounters in, for
instance, solving a cubic for $p^2$. Indeed, because the
characteristic equation is a quartic in $p$, the qualifying roots
are those with a positive (negative) imaginary part to ensure
decay with distance from the interface in the lower (upper)
half-space. Thus, for the lower half-space say, they \textit{must}
be of the form $\alpha_1 + i \beta_1$, $\alpha_2 + i \beta_2$,
where $\beta_1$ and $\beta_2$ are positive real numbers. This
situation is in sharp contrast with the case of a wave propagating
in a symmetry plane; then the characteristic equation is a cubic
in $p^2$ and the relevant roots for the lower half-space can come
in one of two forms: either as $\ii \alpha_1$, $\ii \alpha_2$,
$\ii \alpha_3$, where $\alpha_1$, $\alpha_2$, $\alpha_3$ are
positive, or as $\ii \alpha_1$, $\pm \alpha + i \beta$, where
$\alpha_1$ and $\beta$ are positive. Although the roots of a cubic
are seemingly easier to obtain analytically than those of a
quartic, determining which of these two forms applies is a tricky
matter. Here, once $p_1$, $p_2$ and $p_1^*$, $p_2^*$ are known,
$M(v)$ and $M^*(v)$ can be constructed and the interfacial wave
speed can be computed directly. The analysis below is conducted
for the lower half-space ($x_2 \ge 0$) and indications are given
at the end of the section on how to adapt it to the upper
half-space.

First write the four-component vectors ${\bm \zeta}^{(1)}$ and
${\bm \zeta}^{(2)}$ as \be {\bm \zeta}^{(1)} =
 \begin{bmatrix}
  {\bm a}^{(1)} \\ {\bm b}^{(1)}
 \end{bmatrix},
 \quad
{\bm \zeta}^{(2)} =
 \begin{bmatrix}
  {\bm a}^{(2)} \\ {\bm b}^{(2)}
 \end{bmatrix}.
\la{3.1}
\en
The vectors ${\bm a}^{(1)}$, ${\bm a}^{(2)}$ are
determined from \rr{2.9}, that is from
\be
[p_k^2  \t + p_k (\r + \r^T)
 + \q - \rho v^2 I]{\bm a}^{(k)} = \o,
\quad k=1, 2,
 \la{3.2}
\en
and the vectors ${\bm b}^{(1)}$, ${\bm b}^{(2)}$, are
computed from \rr{2.11} according to
\be
{\bm b}^{(k)} =
 (p_k \t + \r^T) {\bm a}^{(k)},
\quad k=1, 2.
\la{3.3}
\en
It then follows from \rr{2.19} that
\be
A = \left[{\bm a}^{(1)} | {\bm a}^{(2)} \right],
\quad
B = \left[{\bm b}^{(1)} | {\bm b}^{(2)} \right]
  = T A \Omega + R^T A,
\la{3.4a} \en where \be \Omega =
 \begin{bmatrix} p_1 & 0 \\ 0 & p_2 \end{bmatrix}.
\la{Omega} \en The two eigenvalues $p_1$ and $p_2$ are determined
from $\text{det } (N - p I)=0$, or equivalently, from \be
\text{det } [p^2 \t + p (\r + \r^T) + \q - \rho v^2 I] = 0.
\la{3.4} \en This characteristic equation, called the
\textit{propagation condition}, is a quartic which can be written
as \be p^4 + d_3 p^3 + d_2 p^2 + d_1 p + d_0 = 0, \la{3.5} \en
say, where $d_0, d_1, d_2, d_3$ are real constants.

The usual strategy for solving a quartic equation such as
\rr{3.5} is to use a substitution to eliminate the cubic term;
see Bronshtein and Semendyayev \cite{BrSe97}.
Thus, with the substitution $p = q - d_3/4$,
equation \rr{3.5} reduces to
\be
 q^4 + r q^2 + s q + h = 0,
\la{3.6}
\en
where
\be \label{rsh}
 r = d_2 - \textstyle{\frac{3}{8}} d_3^2,
\quad
 s = d_1 - \textstyle{\frac{1}{2}} d_2d_3
      + \textstyle{\frac{1}{8}} d_3^2,
\quad
 h = d_0 - \textstyle{\frac{1}{4}} d_1 d_3
       + \textstyle{\frac{1}{16}} d_2 d_3^2
        - \textstyle{\frac{3}{256}} d_3^4.
\en
The behavior of the roots of \rr{3.6} depends on the cubic resolvent
\be
 z^3 + 2r z^2 + (r^2 - 4 h) z - s^2 = 0.
 \la{3.8}
\en In particular, Eq.\rr{3.6}, and hence Eq.\rr{3.5}, have two
pairs of complex conjugate solutions if and only if \rr{3.8} has
three real roots $z_1$, $z_2$, $z_3$ such that $z_2 < z_3 \le 0
\le z_1$. Then,
\begin{align}
& p_1 = \textstyle{\frac{1}{2}}
         [\text{sign}(s)\sqrt{z_1} + \ii (\sqrt{-z_2} + \sqrt{-z_3})]
           - \dfrac{d_3}{4},
\notag \\
& p_2 = \textstyle{\frac{1}{2}}
          [-\text{sign}(s)\sqrt{z_1} + \ii (\sqrt{-z_2} - \sqrt{-z_3})]
            - \dfrac{d_3}{4},
\la{3.9}
\end{align}
where  $\text{sign}(s)$ equals 1 if $s$ is non-negative and $-1$
otherwise (this definition overrides the standard definition in
which $\text{sign}(0)=0$). The three roots of \rr{3.8} are \be 2
\lambda^{1/3} \cos (\phi)- \textstyle{\frac{2}{3}} r, \quad 2
\lambda^{1/3} \cos (\phi + 2\pi/3) - \textstyle{\frac{2}{3}} r,
\quad 2 \lambda^{1/3} \cos (\phi + 4\pi/3) -
\textstyle{\frac{2}{3}} r, \la{3.10} \en where \be \lambda =
  \textstyle{\frac{1}{27}}
     (12 h + r^2)^{3/2}, \quad
\cos 3 \phi =
  \textstyle{\frac{27}{2}}
     (12h+r^2)^{-3/2}
      (\textstyle{\frac{2}{27}} r^3
          + s^2 - \textstyle{\frac{8}{3}} r h).
\la{3.11}
\en
Without loss of generality it can be assumed
that $0 \le 3 \phi \le \pi$.
It is easy to show that in this interval,
\be
 \cos (\phi + 2 \pi/3) \le
   \cos (\phi+4 \pi/3) \le
     \cos \phi.
\en
It then follows that the three roots of \eqref{3.8} can explicitly be
identified as
\begin{align}
& z_1 = 2 \lambda^{1/3} \cos (\phi)
         - \textstyle{\frac{2}{3}} r,
\nonumber \\
& z_2 = 2 \lambda^{1/3} \cos (\phi + 2 \pi/3)
        - \textstyle{\frac{2}{3}} r,
\nonumber \\
& z_3 = 2 \lambda^{1/3} \cos (\phi + 4 \pi/3)
         - \textstyle{\frac{2}{3}} r.
\la{3.12}
\end{align}
Since $z_2 \ne z_3$, we may further deduce that $\phi \ne 0$ so
that $0<\phi \le \pi/3$.

Now all the ingredients are in place to compute explicitly $p_1$,
$p_2$; then $A$, $B$; and ultimately, $M(v)$, which is all that is
required to find the speed of the Rayleigh wave propagating at the
interface between the lower half-space and a vacuum. To compute
the speed of Stoneley waves and slip waves, $M^*(v)$ is needed.
For the upper half-space $x_2 \le 0$, the analysis above can be
repeated by starring all the quantities involved. Now the
qualifying roots $p_1^*$, $p_2^*$ must have negative imaginary
parts in order to satisfy the decaying condition and so \rr{3.9}
is replaced by
\begin{align}
& p^*_1 = \textstyle{\frac{1}{2}}
  [\text{sign}(s^*)\sqrt{z^*_1} - \ii (\sqrt{-z^*_2} + \sqrt{-z^*_3})]
           - \dfrac{d^*_3}{4},
\notag \\
& p^*_2 = \textstyle{\frac{1}{2}}
 [-\text{sign}(s^*)\sqrt{z^*_1} - \ii (\sqrt{-z^*_2} - \sqrt{-z^*_3})]
            - \dfrac{d^*_3}{4}.
\la{p1*p2*}
\end{align}
As a result, $M^*(v)$ can be obtained from  $M(v)$ by first
replacing the material constants by their starred counterparts and
then taking the complex conjugate.

\section{How to find the wave speed}

This section presents simple algorithms that can be used to
compute directly the surface wave speed and the interfacial wave
speed. They rely on the use of a symbolic manipulation package
such as Mathematica. The algorithm for Rayleigh waves
(solid/vacuum interface) is presented in detail, and then modified
for Stoneley waves (rigid solid/solid interface).

\subsection{Rayleigh waves}

A \textit{Rayleigh wave} satisfies the traction free boundary
condition ${\bm l}(\mathbf{0}) = {\bf 0}$. Thus, from \rr{M}$_1$,
its speed is given by \be \label{Rayleigh} \text{det } M(v) =
 \text{det }(-\ii BA^{-1}) =
   -\text{det } (T A \Omega A^{-1} + R^T)
    = 0.
\en

Given a set of material constants $C_{ijkl}$ and $\rho$, the
unique surface-wave speed is found by using the following robust
numerical procedure:
\begin{enumerate}
\item[(i)] Enter the values of the stiffnesses into the
definitions \rr{2.12} of $T$, $R$, $Q$,
that is
\[
Q = \begin{bmatrix} C_{1111} & C_{1112} \\
                    C_{1112} & C_{1212}
    \end{bmatrix}, \quad
R = \begin{bmatrix} C_{1112} & C_{1122} \\
                    C_{1212} & C_{2212}
    \end{bmatrix}, \quad
T = \begin{bmatrix} C_{1212} & C_{1222} \\
                    C_{1222} & C_{2222}
    \end{bmatrix}.
\]

\item[(ii)] Expand the quartic \rr{3.4} in $p$,
and obtain the coefficients $d_1$, $d_2$, $d_3$ by
comparing it to \rr{3.5}.

\item[(iii)] Enter the coefficients $r$, $s$, $h$ according to
\rr{rsh} and then enter $p_1, p_2$ according to \rr{3.9} and \rr{3.12}.

\item[(iv)] Define ${\bm a}^{(1)}$ and ${\bm a}^{(2)}$
according to \be
 {\bm a}^{(k)} = \begin{bmatrix}
      p_k^2 T_{12}+p_k (R_{12}+R_{21})+Q_{21} \\
     -p_k^2 T_{11}-2 p_k R_{11}-Q_{11}+ \rho v^2
                   \end{bmatrix}, \;\;\;\; k = 1, 2.
\en

\item[(v)] Define $A$ from \rr{3.4a}$_1$, $\Omega$ from
\rr{Omega}.

\item[(vi)] Use \rr{Rayleigh} and the command
{\tt FindRoot} in Mathematica to solve $\text{det } M(v) = 0$.
\end{enumerate}

To facilitate calculations, the term $\text{sign}(s)\sqrt{z_1}$ in
\rr{3.9} may be replaced by $s/\sqrt{z_2 z_3}$. If necessary, the
solution for $M$ can be checked by substituting the result into
the matrix equation \rr{Ricatti}.

\subsection{Stoneley waves}

A \textit{Stoneley wave} must satisfy the continuity  conditions
\[
  {\bm z}(\mathbf{0}) =  {\bm z}^*(\mathbf{0}),
 \quad
   {\bm l}(\mathbf{0}) =  {\bm l}^*(\mathbf{0}).
\]
It then follows from \rr{M} that $[M(v) + M^*(v)]
{\bm z}(\mathbf{0}) =0$ and so the Stoneley wave speed is
determined by \be \label{Stoneley} \text{det } [M(v) + M^*(v)] =
0, \en where \be \label{MM*} M(v) = -i[T A \Omega A^{-1} + R^T],
\quad M^*(v) = i[T^* A^* \Omega^* (A^*)^{-1} + (R^*)^T]. \en

Given two sets of material constants $C_{ijkl}$, $C^*_{ijkl}$ and
$\rho$, $\rho^*$, the unique Stoneley wave speed, if it exists, is
found by the following robust numerical procedure:
\begin{enumerate}
\item[$\bullet$]
Follow steps (i) to (iv) of the algorithm described in the
preceding subsection twice: once for the lower half-space, and
once for the upper half-space, replacing each quantity but $v$ by
their starred counterpart, taking care in step (iii) that $p^*_1$,
$p_2^*$ are defined by \rr{p1*p2*}.
\item[$\bullet$]
Use \rr{Stoneley}, \rr{MM*} and the command
{\tt FindRoot} in Mathematica to find the speed of the Stoneley wave,
when it exists.
If Mathematica is unable to find a root, then the Stoneley wave
does not exist.
\end{enumerate}

\section{Examples}

We now apply the algorithms to specific materials. We use the data
and results of Chevalier et al. \cite{ChLM91}, where $\theta =
\theta^* = 0$, as a guideline. By varying these angles, we find
that there are situations where a Stoneley wave does not exist and
that, when it does exist, it is faster than the slower Rayleigh
wave associated with either of the two half-spaces; these features
were proved for any anisotropic crystal by Barnett et al.
\cite{BLGM85}. We also compute the smallest imaginary part of the
attenuation coefficients, which is $\Im(p_2)$ according to
\rr{3.9}; this quantity is related to the penetration depth of the
interfacial wave into the substrates: the smaller it is, the
deeper is the penetration.

\subsection{(Aluminum)-(Tungsten) bi-material}

For a bi-material made of aluminum above ($x_2 \le 0$) and of
tungsten below ($x_2 \ge 0$), Chevalier et al. \cite{ChLM91}
found a Stoneley wave propagating at speed 2787 m/s when $\theta =
\theta^* =0$. We recovered this result and extended it to the
consideration of the bi-material obtained when the half-spaces are
rotated symmetrically about the $Z$ axis before they are cut and
bonded,
\begin{equation}
  0^\circ \le \theta \le 90^\circ, \quad \theta^* = -\theta.
\end{equation}
\begin{figure}[!t]
\centering
\epsfig{figure=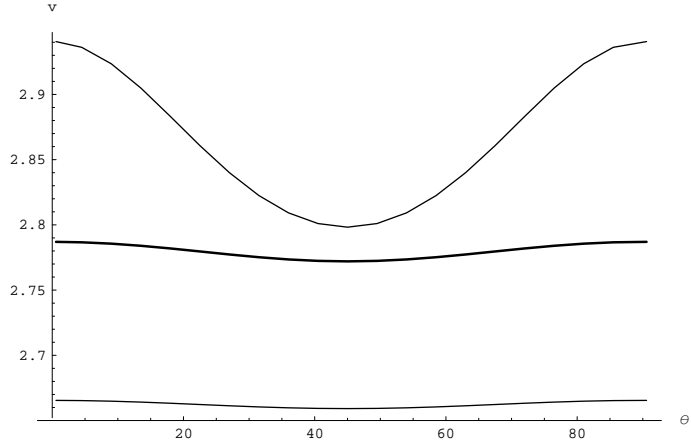,width=.7\textwidth}
 \renewcommand{\figurename}{Fig.}
 \caption{Rayleigh (thin curves) and Stoneley (thick curve) wave
speeds for a bi-material made of aluminum
(rotated $Y$-cut  about the $Z$ axis)
and of tungsten (symmetrically rotated $Y$-cut  about the $Z$ axis).}
\end{figure}

Figure 2 shows that the Stoneley wave (thick curve)
exists for all angles;
it is faster than the Rayleigh wave for a half-space made of
tungsten (lower thin curve) and slower
than the Rayleigh wave for a half-space made of
aluminum (upper thin curve).
\begin{figure}[!t]
\centering
\epsfig{figure=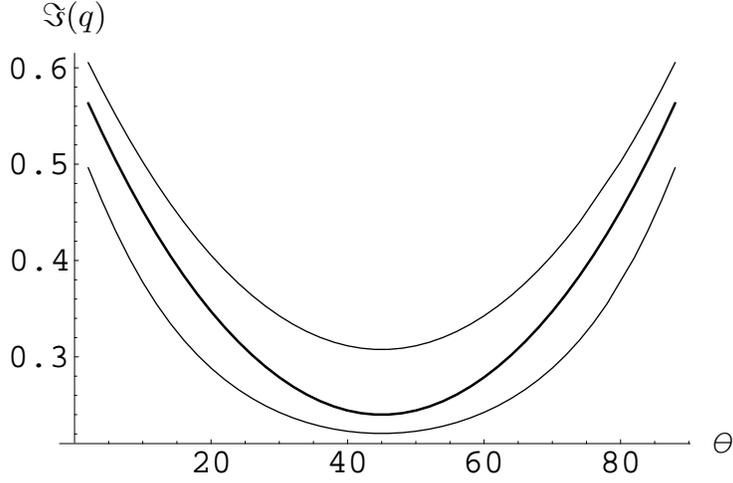,width=.7\textwidth}
 \renewcommand{\figurename}{Fig.}
 \caption{Attenuation factors for Rayleigh (thin curves) and Stoneley
(thick curve) wave
speeds for a bi-material  made of aluminum
(rotated $Y$-cut  about the $Z$ axis)
and of tungsten (symmetrically
rotated $Y$-cut  about the $Z$ axis).}
\end{figure}

Figure 3 displays the smallest imaginary part of the
attenuation factors for each wave.
We find that for the Stoneley wave, this quantity is intermediate
between the corresponding quantities for the Rayleigh waves,
indicating a similar localization.

\subsection{(Douglas pine)-(Carbon/epoxy) bi-material}

For a bi-material made of carbon/epoxy above ($x_2 \le 0$) and of
Douglas pine below ($x_2 \ge 0$), Chevalier
et al. \cite{ChLM91} found a
Stoneley wave propagating at speed 1353.7 m/s when $\theta^* =0$
and $\theta = 90^\circ$.
Here we investigate what happens to this wave when the half-space
below is rotated about the $Z$ axis before it is cut and bonded,
while the half-space above is left untouched,
\begin{equation}
  0^\circ \le \theta \le 90^\circ, \quad \theta^* = 0.
\end{equation}
\begin{figure}[!t]
\centering
\epsfig{figure=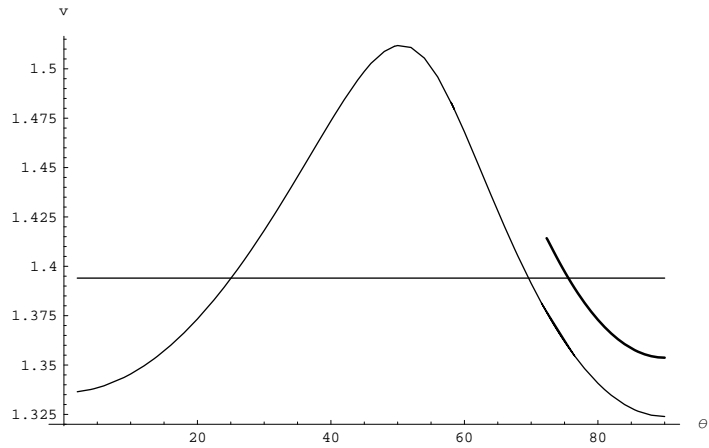,width=.7\textwidth}
 \renewcommand{\figurename}{Fig.}
 \caption{Rayleigh (thin curves) and Stoneley (thick curve) wave
speeds (km/s) for a bi-material made of  Douglas pine ($Y$-cut)
and of  carbon/epoxy (rotated $Y$-cut  about the $Z$ axis).}
\end{figure}

Figure 4 shows that the Stoneley wave indeed exists at $\theta =
90^\circ$ and in the neighborhood of that angle, approximatively
in the range: $72.4^\circ \le \theta \le 107.6^\circ$. In that
range, the Rayleigh wave for a half-space made of carbon/epoxy cut
at an angle $\theta$ (thin varying curve) is slower than the
Rayleigh wave for a half-space made of Douglas pine cut at an
angle $\theta^*=0$ (thin horizontal curve). The Stoneley wave,
when it exists (thick curve), is always faster than the former,
and either faster (for $72.4^\circ < \theta < 75.3^\circ$ and for
$104.7^\circ < \theta < 107.6^\circ$) or slower (for $75.3^\circ <
\theta < 104.7^\circ$) than the latter.
\begin{figure}[!t]
\centering
\epsfig{figure=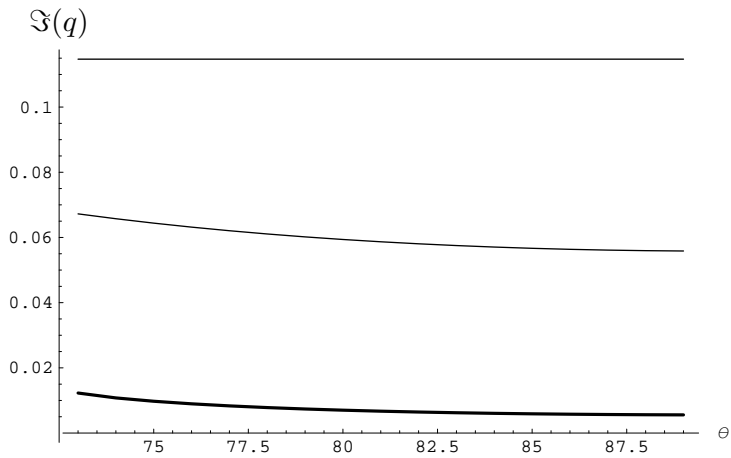,width=.7\textwidth}
 \renewcommand{\figurename}{Fig.}
 \caption{Attenuation factors for Rayleigh (thin curves) and Stoneley
(thick curve) wave
speeds for a bi-material made of  Douglas pine ($Y$-cut)
and of carbon/epoxy (rotated $Y$-cut  about the $Z$ axis),
in the range of common existence.}
\end{figure}

Finally, Figure 5 displays the smallest imaginary part of the
attenuation factors for each wave.
We find that this quantity is for the Stoneley wave
between one sixth and one tenth of that for the Rayleigh waves,
indicating a much deeper penetration.

\section*{Acknowledgments}
This research was made possible by an exchange scheme between
France and the UK. The support of the British Council (UK) and of
the Minist\`ere des Affaires Etrang\`eres (France), as well as the
hospitality of the two Departments involved, are most gratefully
acknowledged.

%++++++++++++++++++++++++++++++++++++++++++++++++++++++
% bibliography
%++++++++++++++++++++++++++++++++++++++++++++++++++++++

%%%%%%%%%%%%%%

\end{document}